# Estimation of the Scatterer Size Distributions in Quantitative Ultrasound Using Constrained Optimization

Noushin Jafarpisheh
Department of Electrical and
Computer Engineering and
PERFORM Center
Concordia University
Montreal, Canada
n_jafarp@encs.concordia.ca

Ivan M. Rosado-Mendez
Department of Medical Physics,
University of Wisconsin-
Madison, Madison, WI, USA
rosadomendez@wisc.edu

Timothy J. Hall
Department of Medical Physics
University of Wisconsin-Madison
WI, USA
tjhall@wisc.edu

Hassan Rivaz
Department of Electrical and
Computer Engineering and
PERFORM Center
Concordia University
Montreal, Canada
hrivaz@ece.concordia.ca

*Abstract*— Quantitative ultrasound (QUS) parameters such as the effective scatterer diameter (ESD) reveal tissue properties by analyzing ultrasound backscattered echo signal. ESD can be attained through parametrizing backscatter coefficient using form factor models. However, reporting a single scatterer size cannot accurately characterize a tissue, particularly when the media contains scattering sources with a broad range of sizes. Here we estimate the probability of contribution of each scatterer size by modeling the measured form factor as a linear combination of form factors from individual sacatterer sizes. We perform the estimation using two novel techniques. In the first technique, we cast scatterer size distribution as an optimization problem, and efficiently solve it using a linear system of equations. In the second technique, we use the solution of this system of equations to constrain the optimization function, and solve the constrained problem. The methods are evaluated in simulated backscattered coefficients using Faran theory. We evaluate the robustness of the proposed techniques by adding Gaussian noise. The results show that both methods can accurately estimate the scatterer size distribution, and that the second method outperforms the first one.

## I. INTRODUCTION

Quantitative ultrasound (QUS) has improved the capabilities of ultrasound imaging by providing quantitative analysis in the frequency-dependent backscattered signal [1]. QUS parameters such as the Effective Scatterer Diameter (ESD) can be obtained by parametrizing backscattered echo signals using form factors to obtain characteristic information about tissue microstructure [2]. Nevertheless, when the scatterer size is too small or too large compared to the wave length, information about scatterer properties is not reachable [3] and the reported value of ESD can be underestimated or overestimated [4]. Therefore, if these conditions are not achieved, it is possible to extract subresolution information from the frequency-dependent analysis of backscattered echo signals.

Prior researchers demonstrated the various clinical applications of ESD estimation such as cancer detection [5], [6], differentiation of normal and tumor tissues [7]–[9], and monitoring disease [10]. To obtain ESD, a Gaussian model can be fit to the backscatter coefficient (BSC) according to the method described in [11]. However, in media with different scatterer size, to obtain ESD, BSC parametrization is not guaranteed to lead to the accurate results [12]. In addition, Lavarello *et al.* [13] reported how the distribution of scatterer size affect the estimation of ESD. Lavarello *et al.* [14] and Nordberg *et al.* [15] illustrated the effect of frequency on estimating ESD especially when the range of scatterer sizes is broad. Subsequently, scatterer size distribution needs to be taken into account and reporting a single size as ESD in a tissue does not tell the whole story about tissue microstructure [12].

Shape and scatterer size distribution affect radio frequency (RF) data [16]. The impact of aggregate size distribution on backscattered data has been investigated in publications such as [17], [18]. Furthermore, analyzing scatterer size distribution can conduct us to deduce the tissue microstructure. For instances, aggregate size distributions can occur as a consequence of the presence of a tumor; [16] thyroid follicles have a wide range of scatterer size distributions [19]. Nordberg *et al.* [15] employed Faran theory and a Gaussian form factor to separately calculate the contribution of each scatterer size and then combined the contributions.

In the present work, we aim at estimating the scatterer size distribution which reveals the contribution of each scatterer size individually to the BSC. The proposed method is validated using four simulated phantoms with various distributions. We observe that the estimated distributions correspond well with the expected results.

The main strength of the proposed methods is that they can accurately estimate the scatterer size distribution. Further, they do not depend on the shape of ground truth ESD distribution. Therefore, the techniques can be very good options for clinical application where the ground truth is unknown.

In Section II, the proposed methods are described, where we present how the scatterer size distribution can be estimated through matrix equations. Results and discussion are provided in Section III. Finally, we conclude the paper in Section IV.

## II. METHODS

### A. Data

Four phantoms of 1.6L in volume of a water-based gel, with 200 grams of glass beads with different scatterer size distribution were simulated. The diameter of glass beads was from 1 to 100 $\mu m$ and the bandwidth of interest was from 3 to 9MHz in steps of 0.1MHz. The distribution of scatterer size assumed to be different for each phantom. Table I presents the properties of the simulated phantoms.

### A. Algorithm

Starting with a bank $F$ of predefined form factors for different scatterer sizes, the resulting form factor $F_T$ for the entire size distribution is a linear combination of the form factors within the bank as follows:

$$F_T = A \times F \quad (1)$$

where $F$ is a $N_a \times N_f$ matrix where $N_a$ is the number of form factors for different scatterer sizes, and $N_f$ is the number of frequency bins within the analysis bandwidth; $A$ is a vector with $N_a$ elements that modulate the contribution of each form factor within $F$, and $F_T$ is a $N_f$ vector corresponding to the total form factor that results from the linear combination of the components of $F$. Thus, the goal is to estimate $A$, which will indicate the contribution of different form factors and, therefore, the contribution of scatterers of different sizes. As $F$ is not necessarily a square matrix, in the first method, $A$ can be obtained using the pseudo-inverse approach as follows:

$$A = F_T \times (F F')^{-1} \quad (2).$$

In the second method, we use the solution of this system of equations to constrain the optimization function. This is achieved by using the results of the first method to suppress the fluctuations and negative values in elements of $A$ observed in the results of the first method.

Both methods are tested on mathematically computed BSC from Faran theory, from which form factors are derived by $f^4$ and renormalizing. Since Faran theory [20] can well represent BSC in tissue mimicking phantoms containing glass beads as the scattering source, we build the bank of form factor using Faran's equation. Nevertheless, we can exploit whatever form factor model and what we only need is to change the bank and precompute it before running the algorithm.

Fig. 1. outlines our proposed algorithm to estimate scatterer size contribution where each sub-index represents $i^{th}$ ($i = 1, …, N_a$) or $j^{th}$ ($j = 1, …, N_F$) element of a vector or a matrix. In the second method, we perform the constrained optimization and zero out the contribution for scatterer sizes which are below the considered threshold. The threshold is selected to cancel the highest absolute fluctuation in elements of $A$. In Fig. 1., we assumed the fluctuation is observed in probability of scatterer size from $k_1$ to $k_t$. Based on our second method, the elements of matrix $F$ correspond to size from $k_1$ to $k_t$ should be replaced by zero. Accordingly, we obtain a new matrix named $F_c$ to calculate the $A_c$. as follows:

$$A_c = F_T \times (F_c F_c')^{-1} \quad (3).$$

At the end, the robustness of the method is tested by adding zero mean Gaussian noise with the variance in the order of $10^{-5}$ to the BSC.

## III. RESULTS AND DISCUSSION

The results of both methods are shown in Fig. 2. Red and blue curves in each sub-figure present the estimated distribution using first and second methods, respectively, the green curves indicate the results after adding Gaussian noise, and the black dash line accounts for expected distributions. It is clearly observed that both methods can estimate different unimodal (Figs. 2(a) and (b)), uniform (Fig. 2(c)) and bimodal (Fig. 2(d)) scatterer size distributions very accurately, and that the second method outperforms the first one. As it is expected, adding white Gaussian noise to the BSC increases the fluctuations in the estimated distribution using the first method, but the second method can still mitigate those variations. It is worth mentioning that, only the scatterer sizes corresponding to $0.6 \leq ka \leq 1.2$ are estimable regardless of the method of estimation, where $k$ is the wave number and $a$ is the scatterer size. Solving this inequality for $f=3$ (lower frequency band) and $f=9$ (upper frequency band) yield $a = 95.5$ and $a = 15.9$ which means only the scatterer sizes from 15.9 $\mu m$ to 95.5$\mu m$ can be estimated within the frequency bandwidth. In addition, the results show that our methods provide almost zero contribution for the scatterer sizes beyond this range, but we need to test if our method can still work when the expected distributions include these values.

This work had various limitations. The fluctuations illustrated in the estimated distributions using the first method emanates from *pinv* built-in function of Matlab. We are currently working on developing methods to overcome this limitation. Additionally, this work needs to satisfy some assumptions that limit its capability, for example, having single and weak scattering.

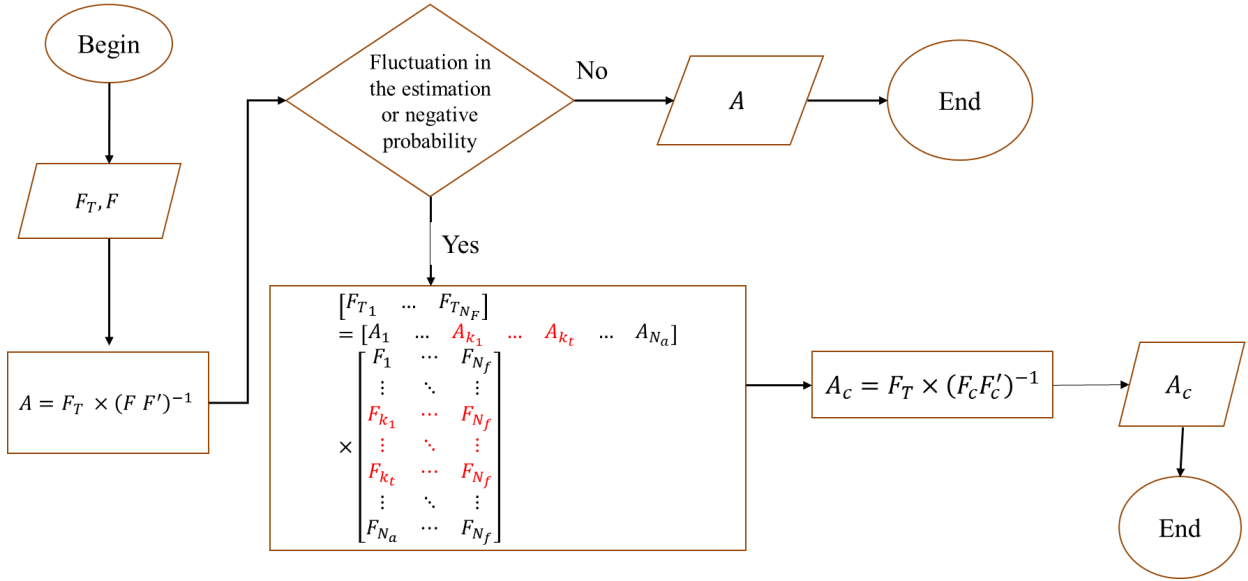

Fig. 1. Algorithm of the proposed methods.

TABLE I
PROPERTIES OF THE SIMULATED PHANTOMS

| | |
|---|---|
| Mass of beads in the phantom | 200 gr |
| Volume of the phantom | 1.6 L |
| Bead diameter | 1-100 $\mu m$ |
| Poisson's ratio for the sphere | 0.210 $\mu m$ |
| Sound speed in the sphere | 5.5719 $\frac{mm}{\mu s}$ |
| Sound speed in the background | 1.498 $\frac{mm}{\mu s}$ |
| Mass density of the sphere | 2.38 $\frac{g}{cm^3}$ |
| Mass density of the background | 1.04 $\frac{g}{cm^3}$ |

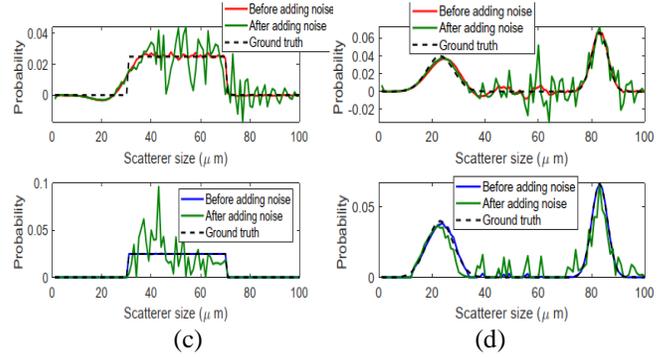

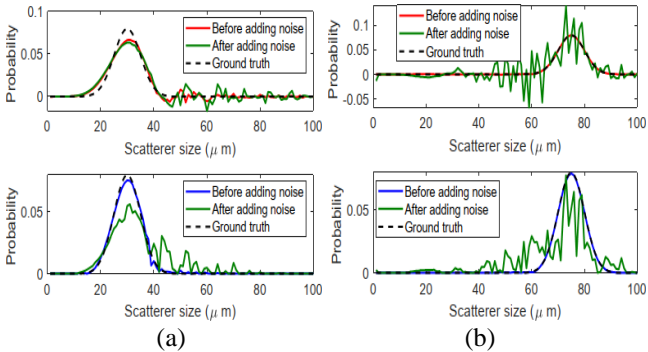

Fig. 2. ESD in four simulated scatterer size distributions (a)-(d). The top and bottom plots in each sub-figure present the estimated distribution using the first and the second methods, respectively.

## IV. CONCLUSION

Tissues are usually composed of combinations of acoustic scatterers with different sizes and what changes during a disease is the distribution of sizes. Thus, the proposed method provides a comprehensive characterization of tissue by quantifying the scatterer size distribution using system of linear equations and evaluating the robustness of our method by adding white Gaussian noise to the BSC. Although in this work we used form factors from Faran theory, different models can be used by changing the bank. The bank can be precomputed before running the algorithm to save processing time. As the next step we intend to implement our techniques to data acquired from tissue mimicking phantoms.

## V. ACKNOWLEDGMENT

We acknowledge the support of the Natural Sciences and Engineering Research Council of Canada (NSERC) RGPIN-2020-04612, and NIH R01HD072077.